\documentclass[11pt]{article}
\usepackage{amsfonts}
\usepackage{amsmath,amssymb,amscd}
\usepackage{stmaryrd}
\usepackage[dvips]{color}
\usepackage[dvipdfmx]{hyperref}
\usepackage{fancybox}
\usepackage{cite}
\usepackage{bm}
\usepackage[all]{xy}
\usepackage{wrapfig}
\usepackage{ulem}   
\usepackage{comment}

\baselineskip=\normalbaselineskip

\begin{document} 

\numberwithin{equation}{section}
\renewcommand{\theequation}{\thesection.\arabic{equation}}
%%%%% Begin the Note, use \section{} %%%%%%%%%
\def\natural{\mathbb{N}}
\def\mat#1{\matt[#1]}
\def\matt[#1,#2,#3,#4]{\left(%
\begin{array}{cc} #1 & #2 \\ #3 & #4 \end{array} \right)}
\def\hq{\hat{q}}
\def\hp{\hat{p}}
\def\hx{\hat{x}}
\def\hk{\hat{k}}
\def\hw{\hat{w}}
\def\hl{\hat{l}}

%%%%%%special for this file only %%%%%%%%%%%%
\def\bea#1\ena{\begin{align}#1\end{align}}
\def\bean#1\enan{\begin{align*}#1\end{align*}}
\def\p{\partial}
\def\nn{\nonumber\\}
\def\cL{{\cal L}}
\def\cE{{\cal E}}
%%% Prepared for Generalized Geometry %%%%%%%%
\def\TM{TM\oplus T^*M}
\newcommand{\CouB}[2]{\left\llbracket #1,#2 \right\rrbracket}
\newcommand{\pair}[2]{\left\langle\, #1, #2\,\right\rangle}

\null \hfill TU-1034, UTHEP-694, KEK-TH-1940  \\[3em]
\begin{center}
{\LARGE \bf{
Contravariant Gravity on Poisson Manifolds \\[0.5em]
and Einstein Gravity
}}\\[2em] 
\end{center}

\begin{center}
{Yukio Kaneko${}^{\sharp}$\footnote{
e-mail: y\_kaneko@tuhep.phys.tohoku.ac.jp}},
{Hisayoshi Muraki${}^{\flat\dagger}$\footnote{
e-mail: hmuraki@het.ph.tsukuba.ac.jp}}
and {Satoshi Watamura${}^{\sharp}$\footnote{
e-mail: watamura@tuhep.phys.tohoku.ac.jp}}\\[2em] 

${}^{\sharp}$ Tohoku University,\\
Graduate School of Science,\\
Aoba-ku, Sendai 980-8578, Japan\\[1em]

${}^{\flat}$ University of Tsukuba,\\
Graduate School of Pure and Applied Sciences,\\
Tsukuba, Ibaraki 305-8571, Japan\\[1em]

${}^{\dagger}$ High Energy Accelerator Research Organization,\\
KEK Theory Center,\\
Tsukuba, Ibaraki 305-0801, Japan\\[3em]

\thispagestyle{empty}

\abstract{
\noindent
A relation between gravity on Poisson manifolds proposed in \cite{AMW}
and Einstein gravity is investigated.
The compatibility of the Poisson and Riemann structures defines a 
unique connection, the contravariant Levi-Civita connection, 
and leads to the idea of the contravariant gravity.
The Einstein-Hilbert-type action yields an equation of motion
which is written in terms of the analog of the Einstein tensor, and it 
includes couplings between the metric and the Poisson tensor.
The study of the Weyl transformation reveals properties of those interactions.
It is argued that this theory
can have an equivalent description as a system of Einstein gravity coupled to matter.
As an example, it is shown that the contravariant gravity on a two-dimensional Poisson manifold
can be described by a real scalar field coupled to the metric in a specific manner.
} 

\end{center}

\vskip 2cm

\eject
%%%%%%%%%%%%%%%%%%%%%%%%%%%%%%%%%%%%%%%%%%%%%%%

 \tableofcontents

%%%%%%%%%%%%%%%%%%%%%%%%%%%%%%%%%%%%%%%%
%%%%%%%%%%%%%%%%%%%%%%%%%%%%%%%%%%%%%%%%
\section{Introduction}

Both Riemann and Poisson structures play significant roles in physics and mathematics.
The former provides us with a geometrical intuition about spacetime and gravity,
while the latter provides us with a geometrical intuition about the 
time evolution of a system and quantization.
Here, we investigate the interplay between these two structures
through a consideration on gravity on Poisson manifolds.

In quantum mechanics, it is well-known that 
a point in phase space can
only be determined to an accuracy of the order of Planck's constant $\hbar$. 
The Heisenberg uncertainty principle can be regarded as a measure of the noncommutativity in the quantum world.
The Poisson structure in phase space appears as a semi-classical approximation of the quantized system.
Analogously, a Poisson structure of spacetime can be interpreted as a semi-classical approximation of the spatial noncommutativity.
The idea of noncommutative spacetime can be traced back to Snyder \cite{Snyder}.
Since then the application of this idea
to physics, especially to gravity,
has been discussed intensively in the literature \cite{BGGK,HR}.
For a review see for example \cite{DN,Szabo}.

Recently, in the string theory context
the concept of a minimal length is discussed.
It is argued that it can be understood as an origin of noncommutativity in probing spacetime geometry \cite{Yoneya,Yoneya1}.
For review see for example \cite{SH}. 
There is also a proposal of gravity on noncommutative spacetime in relation with deformation quantization \cite{JW,J}. It is based on the quantization 
of the diffeomorphism by twisting the Hopf algebra structure and there, a noncommutative algebra is formulated by the star product. See also \cite{KS}.

One possible scenario would be that a gravity theory on noncommutative spaces
gives an effective theory of quantum gravity.
The noncommutativity should be very small and
the effective theory can be described well by a semi-classical approximation.
We can imagine that the relic of noncommutativity 
appears as a Poisson structure, and that the resulting Poisson 
structure should be compatible with the Riemann structure.
From this point of view, gravity on noncommutative spaces and
its semi-classical approximation, i.e. a gravity on Poisson manifolds, should be
issues worthy of being studied.

The gravity on Poisson manifolds, 
which we refer here to as contravariant gravity,
is formulated by Riemannian geometry 
compatible with a Poisson structure.
The corresponding geometry  
is specified by a unique connection
consisting of a Riemannian metric and a Poisson tensor,
called the contravariant Levi-Civita connection.
 Such a geometry has been advocated in physics \cite{AMW,AMSW} as well as in 
mathematics \cite{fernandes2000,Boucetta:2011ofa2,Boucetta:2011ofa3,KSmodular,Hawkins:2002rf}.
This geometry enables us to introduce the Poisson tensor as
an intrinsic geometrical degree of freedom rather than a matter degree of freedom.

It is an interesting question to ask 
how the effect of the Poisson tensor incorporated in the geometry of 
contravariant gravity looks like.
Can the effect be represented as a matter field in the usual Einstein gravity?
Similar considerations  are also found, for example,  in Kaluza-Klein theory, where
a five-dimensional metric governed by the Einstein-Hilbert action
is decomposed into a four-dimensional metric, a $U(1)$ gauge field and a real scalar field,
forming the four-dimensional Einstein-Maxwell dilaton theory.
The aim of this article is to shed some light on these issues.

The organization of this article is as follows. In section 2, we give a brief review on
the Riemannian geometry relevant for the contravariant gravity. 
The invariant measure and the divergence theorem are also discussed. 
In section 3 we give an action which describes  
the Einstein-Hilbert action in such a geometry.
This theory,
i.e. the contravariant gravity, incorporates the Poisson tensor as
a geometrical degree of freedom on an equal footing with the metric,
rather than a matter degree of freedom.
Then, we discuss the equation of motion and
a field redefinition given by a Weyl transformation.
In section 4 we give a concrete example by applying the field redefinition to
the action of our theory to the case of a two-dimensional Poisson-Riemannian manifold.
It is shown that the theory has another description by Einstein gravity
coupling to a real scalar field in a specific manner.
The final section is devoted to summary and discussions.
Appendices are for reference to the reader,
making a brief introduction to the Lie algebroid induced by a Poisson structure
and the Riemannian geometry based on the algebroid,
and presenting some computational details.

%%%%%%%%%%%%%%%%%%%%%%%%%%%%%%%%%%%%
\section{Overview of Riemannian Geometry on Poisson Manifolds}%%%%%%%%%%%

In this section we introduce a geometry describing gravity on a Poisson manifolds,
which we call ``contravariant gravity".
This geometry is based on
a unique connection specified by the metric-compatibility and the torsion-free conditions
\cite{AMW,AMSW}.
See also
\cite{fernandes2000,Boucetta:2011ofa2,Boucetta:2011ofa3,KSmodular,Hawkins:2002rf}.
We give a summary on the definitions of geometrical notions as well as our notations
in appendix \ref{ApenDA}.

%%%%%%%%%%%%%%%%%%%%%%%%%%%%%%%%%%%%
\subsection{Contravariant Derivatives, Torsion and Curvature}%%%%%%%%%%%

Let $M$ be an $n$-dimensional Poisson-Riemannian manifold equipped with a Poisson tensor
$\theta\in \Gamma(\wedge^2 T M)$
 and a Riemannian metric $G \in \Gamma({\otimes^2_{\rm Sym}} (T M))$.
The contravariant Levi-Civita connection
$\bar\nabla: \Gamma(T^*M) \times \Gamma(T^*M) \to \Gamma(T^*M)$
 is defined by
\bea
	\bar\nabla_{dx^i}dx^j = \bar{\Gamma}^{ij}_k dx^k,
\ena
where
\bea
	\bar{\Gamma}^{ij}_k
	&=\frac{1}{2}G_{mk}\left(\theta^{il} \p_l G^{jm}+\theta^{jl} \p_l G^{im}-\theta^{ml} \p_l G^{ij}
	+G^{lj}\p_l \theta^{mi}+G^{li}\p_l \theta^{mj}+G^{lm}\p_l \theta^{ij}\right),	\label{220A}
\ena
for $\theta^{ij}=\theta(dx^i,dx^j)$ and $G^{ij}=G(dx^i,dx^j)$ on a local patch $\{x^i\}$.
For any one-forms $\xi =\xi_idx^i$ and $\eta =\eta_idx^i$, 
the definition of the connection \eqref{affine} gives the following expression
\bea
	\bar{\nabla}_{\xi} \eta
	&= \xi_i (\theta^{ij} \p_j \eta_k 
	+\bar{\Gamma}^{ij}_k \eta_j )dx^k .	
\ena
In general, the contravariant derivative $\bar{\nabla}_\xi$
for an $(r,s)$-tensor field is given by
\bea
	\bar{\nabla}_\xi T^{i_1 \cdots i_r}_{j_1 \cdots j_s}
	&=\xi_k \left( \theta^{kl}\p_l T^{i_1 \cdots i_r}_{j_1 \cdots j_s}
	+\sum_{q=1}^s \bar{\Gamma}^{kl}_{j_q} 
	 T^{i_1 \cdots i_r}_{j_1 \cdots l \cdots j_s}
	-\sum_{p=1}^r \bar{\Gamma}^{ki_p}_l 
	T^{i_1 \cdots l \cdots i_r}_{j_1 \cdots j_s}  \right) .	\label{CDer}
\ena
The contravariant Levi-Civita connection \eqref{220A}
is specified as the unique solution compatible with the metricity 
$\bar{\nabla}_{dx^k}G^{ij}=0$ and 
the torsion-free condition:
\bea
	&\bar{T}(\xi,\eta):=
		\bar\nabla_{\xi} \eta -\bar\nabla_{\eta} \xi - [\xi,\eta]_\theta=0.
\ena
It also satisfies
\bea
	&\bar\cL_{\xi} G(\eta,\zeta)
		=G (\bar{\nabla}_\xi \eta, \zeta) +G (\eta, \bar{\nabla}_\xi \zeta) .
\ena
Here $\bar\cL$ stands for the contravariant Lie derivative, acting on a function $f$ as
\bea
	\bar\cL_{dx^i} f = \theta^{ij}\p_j f ,	\label{CLieScal}
\ena
and $[\ ,\ ]_\theta$ denotes the Koszul bracket,
which is defined for one-forms $\xi=\xi_{i}dx^{i}$ and $\eta=\eta_{i}dx^{i}$ as
\begin{align}
	[\xi,\eta]_{\theta}
	=\left(\xi_{k}\theta^{kl}\partial_{l}\eta_{i}
	-(\theta^{lk}\partial_{k}\xi_{i}
	+\partial_{i}\theta^{lk}\xi_{k})\eta_{l} \right)dx^{i}\,
	(\equiv\bar{\cL}_\xi\eta) .
\end{align}
The contravariant Lie derivative acting on a tensor of type $(r,s)$ is defined by
\bea
	\bar{\cL}_\xi T^{i_1 \cdots i_r}_{j_1 \cdots j_s}
	&=\xi_k \theta^{kl}\p_l T^{i_1 \cdots i_r}_{j_1 \cdots j_s}
	-\sum_{q=1}^s (M_\xi)^{l}_{\ j_q}  T^{i_1 \cdots i_r}_{j_1 \cdots l \cdots j_s}
	+\sum_{p=1}^r (M_\xi)^{i_p}_{\ l} T^{i_1 \cdots l \cdots i_r}_{j_1 \cdots j_s}  ,
		\label{CovLieT}
\ena
where we have introduced a matrix
\bea
	(M_\xi)^{i}_{\ j}=\theta^{ik}\p_k\xi_j  +  \p_j\theta^{ik} \xi_k.
\ena
For a derivation of these formulas, see reference \cite{AMW}.

The curvature tensor is defined by
\bea
	\bar{R}^{kij}_{l} dx^l :
	=&  (\bar{\nabla}_{dx^i} \bar{\nabla}_{dx^j} -\bar{\nabla}_{dx^j} \bar{\nabla}_{dx^i} 
					- \bar \nabla_{[dx^i,dx^j]_\theta} )dx^k \\
	=& (\theta^{im} \p_m \bar{\Gamma}^{jk}_l 
		- \theta^{jm} \p_m \bar{\Gamma}^{ik}_l -\p_n \theta^{ij} \bar{\Gamma}^{nk}_l 
		+ \bar{\Gamma}^{jk}_m \bar{\Gamma}^{im}_l 
		 - \bar{\Gamma}^{ik}_m \bar{\Gamma}^{jm}_l ) dx^l .	
\ena
The corresponding Ricci tensor is defined
by contracting an upper index of the curvature tensor 
with a lower one, $\bar{R}^{kj}:=\bar{R}^{klj}_l$.
Then, the scalar curvature
is obtained by taking a contraction between the metric  $G$ and the Ricci tensor:
\bea
\bar{R}:=G_{ij}\bar{R}^{ij}=G^{ij}\bar{R}_{ij},\label{Scalarcurvature}
\ena
where $G_{ik}G^{kj}=\delta^j_i$ and $\bar{R}_{ij} = G_{ik}G_{jl}\bar{R}^{kl}$.
The curvature satisfies the following Bianchi identities
\bea
	&\bar{R}^{kij}_{l}+\bar{R}^{ijk}_{l}+\bar{R}^{jki}_{l}=0,		\label{BiA11}\\
	&\bar\nabla_{dx^k} \bar{R}^{mij}_{l} +\bar\nabla_{dx^i} \bar{R}^{mjk}_{l}
	+\bar\nabla_{dx^j} \bar{R}^{mki}_{l}=0. 	\label{BiA21}
\ena
In addition to these identities, there is an additional identity
\bea
	&\bar{R}^{mkij} = -\bar{R}^{kmij}.
\ena
This property, together with the Bianchi identity, guarantees that the Ricci tensor
is a symmetric tensor.
We can also define an analog of the Einstein tensor of the form
\bea
	\bar{\mathcal{G}}^{ij} = \bar{R}^{ij} - \frac{1}{2}G^{ij}\bar{R},	\label{EinT}
\ena
which satisfies
\bea
	\bar{\nabla}_{dx^k} G_{ki}\bar{\mathcal{G}}^{ij}=0,
\ena
owing to the Bianchi identities \eqref{BiA11} and \eqref{BiA21}.

%%%%%%%%%%%%%%%%%%%%%%%%%%%%%%%%%%%%
\subsection{Transformation Laws of $G^{ij}$ and $G=\det G^{ij}$}%%%%%%%%%%%

The contravariant Lie derivative \eqref{CovLieT} of the metric tensor $G^{ij}$ can
 be written in terms of the contravariant derivatives \eqref{CDer} as
\bea
	\bar{\cL}_\xi G^{kl}
	&=\bar{\nabla}_{dx^k}(\xi_i G^{il}) + \bar{\nabla}_{dx^l}(\xi_i G^{ki}),
\ena
where we used a fact that $\bar{\nabla}_{dx^k}G^{ij}=0$.
This is a contravariant counterpart of $\cL_X g_{ij}= \nabla_iX_j+\nabla_jX_i$ in the usual Riemannian geometry.
It is also shown that the Kronecker delta is invariant under the contravariant Lie derivatives:
$\bar{\cL}_\xi  \delta^{k}_{l}=0$. Then we obtain
\bea
	\bar{\cL}_\xi G_{ij} = - G_{ik} (\bar{\cL}_\xi G^{kl}) G_{lj}.
\ena
Introducing $G^{-1}=(\det G^{ij})^{-1}=\det G_{ij}$, we find
\bea
	\bar{\cL}_\xi \sqrt{G^{-1}}
	=\frac{1}{2}\sqrt{G^{-1}} G^{ij} \bar{\cL}_\xi G_{ij}
	=-\sqrt{G^{-1}}  \bar{\nabla}_{dx^i} \xi_i .	\label{Mahler}
\ena
Using an identity
\bea
	\sqrt{G^{-1}}\bar{\Gamma}^{ji}_j
	&=-  \p_j (\sqrt{G^{-1}} \theta^{ij} ),	\label{debussy}
\ena
we find a useful relation:
\bea
	\sqrt{G^{-1}}  \bar{\nabla}_{dx^i} \xi_i 
	&=\p_j (\sqrt{G^{-1}}\theta^{ij} \xi_i ) +2\sqrt{G^{-1}}\bar{\Gamma}^{ji}_j\xi_i.	\label{Totd}
\ena
There is a crucial difference between the above relation and the one in usual Riemannian geometry: 
The covariant counterpart is given by 
\bea
	\sqrt{g}\nabla_i X^i=\p_i(\sqrt{g}X^i),	\label{CovLTD}
\ena
which plays an important role in the proof of the general covariance of the action integral.
On the other hand, in the contravariant geometry the quantity 
$\bar{\Gamma}^{ji}_j$ in \eqref{Totd} arises as an obstruction to 
the ``covariance'' of
the na\"ive integral measure $\sqrt{G^{-1}}d^nx$.
This obstruction forces us to introduce an additional factor into the invariant measure.
For later purpose, we note the following relation: For any scalar $F$, we have
\bea
	\bar{\cL}_\xi (F \sqrt{G^{-1}} )
	&=\p_j (F\sqrt{G^{-1}}\xi_i \theta^{ij})  +F\sqrt{G^{-1}} \theta^{ij}(\p_i\xi_j -\p_j\xi_i),
		\label{Puccin}
\ena
which is shown by using \eqref{Mahler} and \eqref{debussy}.

%%%%%%%%%%%%%%%%%%%%%%%%%%%%%%%%%%%%
\subsection{Divergence Theorem and Invariant Measure}%%%%%%%%%%%

As discussed above,
a na\"ive integral measure $\sqrt{G^{-1}}$ fails to give a measure invariant under 
the contravariant Lie derivatives.
The invariance is broken by the existence of $\bar{\Gamma}^{ji}_j$ in \eqref{Totd}.
The removal of  this obstruction has already been discussed in \cite{AMW}, where
 an invariant measure is constructed by multiplying $\sqrt{G^{-1}}$ by a scalar factor $e^\phi$.
However, we conclude that it is not possible to obtain an integral which is fully invariant
under the contravariant Lie derivatives, due to \eqref{Puccin}.
We give details on this issue in the latter half of this subsection.

\paragraph{Divergence Theorem}

As in usual Riemannian geometry, we can also formulate the divergence theorem (\ref{CovLTD}) in contravariant  differential calculus. 
An invariant measure $e^\phi\sqrt{G^{-1}}$ satisfies the ``divergence theorem" if it holds
\bea
	e^\phi\sqrt{G^{-1}}  \bar{\nabla}_{dx^i} \xi_i 
	&=\p_j (e^\phi\sqrt{G^{-1}}\xi_i \theta^{ij}  ).	\label{Beeth}
\ena
This requirement imposes a condition on $\phi$:
\bea
	\theta^{ij} \p_j \phi
	&=2\bar{\Gamma}^{ji}_j 
	= - \frac{2}{\sqrt{G^{-1}}} \p_j (\sqrt{G^{-1}} \theta^{ij} ).	\label{Dil}
\ena
The proof is as follows:  
Take the derivation in right-hand side of (\ref{Beeth}) then substitute (\ref{Dil}) 
for $\theta^{ij}\p_j\phi$. Using the formula of $\bar\Gamma^{ji}_j$, we obtain the left-hand side. 

Note that $\bar{\Gamma}=\bar{\Gamma}^{ij}_i \p_j$ defines a vector field 
(see appendix \ref{CaffineC}).
By a straightforward computation, this vector $\bar{\Gamma}$ turns out to be $d_\theta$-closed.
Thus, it makes sense to ask whether the vector $\bar{\Gamma}$ 
is $d_\theta$-exact or not.
The condition \eqref{Dil} is nothing but $d_\theta$-exactness of $\bar{\Gamma}$.
In this article, we assume that the vector field $\bar{\Gamma}$
is a trivial element of the first $d_\theta$-cohomology,
i.e., there exists a function $\phi$ satisfying the equation \eqref{Dil}.

In the case where the Poisson tensor is invertible, i.e. there exists a symplectic structure,
the equation \eqref{Dil} can be solved by the ansatz
\bea
	\phi = - \log \left(\det(\theta^{ij}) \det (G_{kl})\right).
\ena
It is straightforward to show that this ansatz solves the condition \eqref{Dil}.

\paragraph{Invariant Measure}

Here we discuss about the measure which is invariant 
under the transformation generated by the contravariant Lie derivatives. We see that  
it is impossible to obtain an integral which has full invariance. 
It turns out that we have to restrict the transformation 
by imposing a condition on the transformation parameter.

Let us first consider the general condition
for the invariant measure: 
\bea
	\bar{\cL}_\xi (e^\phi \sqrt{G^{-1}} )
	&=C \, \p_j (e^\phi\sqrt{G^{-1}}\xi_i \theta^{ij}  ).	\label{Brahm}
\ena
The above requirement guarantees the invariance of the measure $e^\phi\sqrt{G^{-1}}$ itself\footnote{In \cite{AMW} $C=-1$ is realized for any one-form $\xi$, by
setting $\theta^{ij} \p_j \phi  =  \bar{\Gamma}^{ji}_j$.
This condition for $\phi$ can be re-expressed as $\p_j(e^\phi \sqrt{G^{-1}}\theta^{ij})=0$.}.
In order to define an invariant action, we have also to require invariance of an integration with a scalar function. The condition \eqref{Brahm} gives the relation for any scalar integrand $F$
\bea
	\bar{\cL}_\xi (e^\phi \sqrt{G^{-1}} F)
	&=\p_j (e^\phi \sqrt{G^{-1}}\xi_i \theta^{ij}F)+(C-1) \p_j (e^\phi\sqrt{G^{-1}}\xi_i \theta^{ij}  ) F.
\ena
Hence, we have to require $C=1$\,\footnote{One may impose $\p_j (e^\phi\sqrt{G^{-1}}\xi_i \theta^{ij}  )=0$.
But this condition involves a one-form $\xi$ and the argument is carried out
in a parallel manner with that in the main body.
This is reduced to the case when we set
 $\theta^{ij} \p_j \phi  = \bar{\Gamma}^{ji}_j$.}.

However, the condition (\ref{Brahm}) with $C=1$
 can not be realized for a general one-form $\xi$,
 since the second term in \eqref{Puccin}
 can not be canceled out by any choice of $\phi$. 
 Therefore, in order to satisfy 
 \eqref{Brahm}, the second term 
on the right-hand side in equation \eqref{Puccin} should vanish.
This means that
 we have to restrict the transformation to the covariant Lie derivatives 
induced by the one-form $\xi$ satisfying 
\bea
\theta(d\xi)=\theta^{ij}(\p_i\xi_j -\p_j\xi_i)=0.
\ena
When the parameter is restricted by this condition, from \eqref{Puccin}, 
we automatically obtain \eqref{Brahm} with $C=1$.

%%%%%%%%%%%%%%%%%%%%%%%%%%%%%%%%%%%%
\section{Gravity on Poisson Manifolds}%%%%%%%%%%%%%%%%

In this section, we investigate a gravity theory 
based on the contravariant Levi-Civita connection. 
We investigate the analog of the Einstein-Hilbert action defined by the scalar curvature 
(\ref{Scalarcurvature}) \cite{AMW}. 
First we derive the corresponding equation of motion 
and show that it is given by an analog of the Einstein tensor.
Then, we perform the Weyl transformation for this Einstein-Hilbert action
and discuss how the Poisson tensor couples to the gravity.

%%%%%%%%%%%%%%%%%%%%%%%%%%%%%%%%%%%%
\subsection{Contravariant Gravity Theory}%%%%%%%%%%%%%%%

As we discussed in the previous section,
in order to define the invariant action with the scalar curvature, we restrict the transformation 
parameter $1$-form $\xi$ as 
$\theta(d\xi)=0$. 
Furthermore, we take the measure
satisfying the divergence theorem (\ref{Beeth}). Thus, the measure is defined with the factor $e^{\phi}$ defined by the condition (\ref{Dil}).

%Therefore, t
The action of contravariant gravity is defined by
\bea
	S=\int d^nx e^{\phi}\sqrt{G^{-1}}\bar{R} , \label{EinHilb}
\ena
where 
\bea
	\theta^{ij} \p_j \phi
	= - \frac{2}{\sqrt{G^{-1}}} \p_j (\sqrt{G^{-1}} \theta^{ij} ).	
\ena
We note that a variation of the scalar function $\phi=\phi[\theta,G]$
must also be taken into account when we take the variation of the metric to obtain the equation of motion.

Varying the Lagrangian density with respect to the metric, we find
\bea
	&\delta (e^{\phi}\sqrt{G^{-1}}\bar{R}) 
	=
	e^{\phi}\sqrt{G^{-1}}G_{ij}\delta\bar{R}^{ij} +e^{\phi}\sqrt{G^{-1}}\delta G_{ij}\bar{R}^{ij} 
	+ e^{\phi}\delta\sqrt{G^{-1}}\bar{R}+\delta \phi e^{\phi}\sqrt{G^{-1}}\bar{R} .
\ena
Since the scalar $\phi$
is a solution of the partial differential equation \eqref{Dil},
we must also consider its variation $\delta \phi$:
\bea
	\delta \phi=\phi[\theta,G+\delta G] -\phi[\theta,G] .
\ena 
This implies
\bea
	\theta^{ij} \p_i \delta \phi 
	&= \theta^{ij} \p_i  (\phi[\theta,G+\delta G]-\phi[\theta,G])\nn
	&= -2 \theta^{ij} \p_i \delta(\log \sqrt{G^{-1}})  .
\ena
We find that a solution is given by
\bea
	\delta \phi = -2 \delta(\log \sqrt{G^{-1}}) =  \frac{-2}{ \sqrt{G^{-1}}} \delta  \sqrt{G^{-1}}.\label{eq:dphi}
\ena

To derive the variation of the Ricci tensor, we first examine the variation of the Riemann tensor
$\delta\bar{R}^{kij}_{l}$.
As mentioned in appendix \ref{CaffineC}, the difference
of the connection coefficients $\delta \bar{\Gamma}^{ij}_k 
	= \bar{\Gamma}^{ij}_k[G+\delta G,\theta] - \bar{\Gamma}^{ij}_k[G,\theta]$
is a tensor, and thus, its contravariant derivative
is well-defined:
\bea
	 \bar{\nabla}_{dx^i} \delta \bar{\Gamma}^{jk}_l
	 =& \theta^{im}\p_m   \delta \bar{\Gamma}^{jk}_l +  \delta \bar{\Gamma}^{jk}_m  \bar{\Gamma}^{im}_l
	 		- \delta \bar{\Gamma}^{mk}_l   \bar{\Gamma}^{ij}_m 
			-  \delta \bar{\Gamma}^{jm}_l \bar{\Gamma}^{ik}_m .
\ena
We find
\bea
	 \delta\bar{R}^{kij}_{l} 
	&= \bar{\nabla}_{dx^i} \delta \bar{\Gamma}^{jk}_l
		- \bar{\nabla}_{dx^j} \delta \bar{\Gamma}^{ik}_l ,
\ena
where we used $ \bar{\Gamma}^{ij}_m- \bar{\Gamma}^{ji}_m=\p_m\theta^{ij}$.
Using the above relation, the variation of the Ricci tensor is obtained by taking a contraction
with the metric. 
Using \eqref{Totd} and the definition of $\phi$,  \eqref{Dil}, we obtain
\bea
	&e^{\phi}\sqrt{G^{-1}}G_{ij}\delta \bar{R}^{ij}
		=\p_m[e^{\phi}\theta^{lm}\sqrt{G^{-1}}(G_{ij} \delta \bar{\Gamma}^{ji}_l
		-G_{il}\delta \bar{\Gamma}^{ji}_j)],
\ena
where we  used the divergence theorem \eqref{Beeth}.
As a result, the variation of the action with respect to the metric reads
\bea
	\delta S 
	&=\int d^nx  \bigg\{e^{\phi}\sqrt{G^{-1}}\bigg( \bar{R}^{ij} 
	- \frac{1}{2}G^{ij}\bar{R}\bigg)\delta G_{ij} 
	 +\p[\dots] \bigg\} .
\ena
Thus, the equation of motion of the metric $G_{ij}$ is 
\bea
	\bar{R}^{ij} - \frac{1}{2}G^{ij}\bar{R} =0.	\label{eom1}
\ena
The left-hand side is nothing but the analog of the Einstein tensor 
$\bar{\mathcal{G}}^{ij}$  
defined in (\ref{EinT}).

Let us comment on
the possible terms one may add  
to the Einstein-Hilbert action. As usual, 
one can add a cosmological constant term
\bea
	S_c = 2\Lambda \int d^nx  e^{\phi}\sqrt{G^{-1}}.\label{eq:cc}
\ena
Using the equation (\ref{eq:dphi}), we find that the variation of the cosmological constant term gives
\bea
	\delta S_{c}
	&=2\Lambda\int d^{n}x(\delta\phi e^{\phi}\sqrt{G^{-1}}+e^{\phi}\delta	\sqrt{G^{-1}})\\
	&=-2\Lambda\int d^{n}xe^{\phi}\delta\sqrt{G^{-1}}
	=-\Lambda\int d^{n}xe^{\phi}\sqrt{G^{-1}}G^{ij}\delta G_{ij}.
\ena
It contributes to the equation of motion as follows
\bea
	\bar{R}^{ij} - \frac{1}{2}G^{ij}\bar{R} =\Lambda G^{ij}.	\label{eomm}
\ena

In principle, we can add any diffeomorphism invariant terms 
to the Einstein-Hilbert-type action, such as $\bar{R}^2$, the Weyl tensor etc.
It is also interesting to discuss an analog 
of the Gibbons-Hawking term when a Poisson manifold has boundaries.
In the following section, we mainly focus on the action of the Einstein-Hilbert term \eqref{EinHilb}.

%%%%%%%%%%%%%%%%%%%%%%%%%%%%%%%%%%
\subsection{Weyl Transformation}%%%%%%%%%%%%%%%%

We have introduced a scalar degree of freedom $\phi$ to keep
the divergence theorem \eqref{Beeth}, 
and derived the Einstein equation \eqref{eom1}.
However, its physical interpretation is less clear.
Na\"ively, one might think that it originates from a dilaton field.
In this subsection, we consider the Weyl transformation
and provide transformation rules for the Riemann tensor 
and the scalar field $\phi$.
We also show that an analog of the Weyl tensor can be defined.

Let us consider the Weyl transformation given by
\bea
	&G^{ij} \  \longrightarrow \ \tilde{G}^{ij}= e^{2\Omega} G^{ij}  ,	\label{Weyltra}
\ena
where $\Omega$ is an arbitrary function on the manifold.
We assume that the Poisson tensor does not change under the Weyl transformation.
The coefficients ${\bar{\Gamma}}^{ij}_k$
of the contravariant Levi-Civita connection are transformed as 
\bea
	\tilde{\bar{\Gamma}}^{ij}_k
	=&  \bar{\Gamma}^{ij}_k+
	{G}_{mk}	\big({G}^{jm} \theta^{il}  
			+{G}^{im} \theta^{jl}  -{G}^{ij} \theta^{ml}  \big) \p_l \Omega.
\ena
The curvature tensor becomes 
\bea
	 \tilde{\bar{R}}^{kij}_{l} 
	&={\bar{R}}^{kij}_{l} + {\bar{L}}^{kij}_{l},\label{eq:appppB4}
\ena
where
\bea
	&{\bar{L}}^{kij}_{l} 
	=\delta^{j}_{l}B^{ik}-\delta^{i}_{l}B^{jk}-G^{jk}G_{lm}B^{im}+G^{ik}G_{lm}B^{jm},\\
	&B^{ik}=
	\bar{\nabla}_{dx^{i}}(\theta^{kj}\partial_{j}\Omega)
	-(\theta^{ij}\partial_{j}\Omega)(\theta^{kl}\partial_{l}\Omega)
	+\frac{1}{2}G^{ik}G_{jl}(\theta^{jm}\partial_{m}\Omega)(\theta^{ln}\partial_{n}\Omega).
\ena
For $n$-dimensional manifolds, 
the Weyl transformations of the Ricci tensor and the corresponding Ricci scalar are
\bea
&\tilde{\bar{R}}^{ij}=\bar{R}^{ij}-(n-2)B^{ji}-G^{ij}G_{kl}B^{kl},\label{eq:appppB1}\\
&e^{2\Omega}\tilde{\bar{R}}=\bar{R}-2(n-1)G_{ij}B^{ij}.\label{eq:appppB2}
\ena
Here, we consider the Weyl transformation of the Einstein-Hilbert action. We note that 
under the Weyl transformation the scalar field $\phi$ transforms 
\bea
	\phi \rightarrow\tilde{\phi} = {\phi} + 2n\Omega,
\ena
since the scalar field $\phi$ solves the partial differential equation \eqref{Dil}. It yields
\bea
	e^{\tilde{\phi}}\sqrt{{\tilde{G}}^{-1}} \tilde{\bar{R}}
	=e^{(n-2)\Omega}e^{\phi}\sqrt{G^{-1}} \big( {\bar{R}} + {G}_{kj}{\bar{L}}^{kij}_{i} \big) .
	\label{WLagg}
\ena
Therefore, we can 
eliminate 
the scalar field $\phi$, which appears in the prefactor of the Einstein-Hilbert action,
by performing the Weyl transformation.
We will show that the two-dimensional contravariant gravity theory 
can be rephrased in terms of the usual Einstein-Hilbert action.

It is notable that we also find an analog of the Weyl tensor of the form
\begin{align}
	C^{kij}_{l}=
	&\bar{R}^{kij}_{l}
	+\frac{1}{n-2}(\delta^{j}_{l}\bar{R}^{ki}
	-\delta^{i}_{l}\bar{R}^{kj}-G^{jk}G_{lm}\bar{R}^{mi}+G^{ik}G_{lm}\bar{R}^{mj})\nonumber\\
	&+\frac{1}{(n-1)(n-2)}(\delta^{j}_{l}G^{ik}-\delta^{i}_{l}G^{jk})\bar{R}.
\end{align}
We can check the Weyl invariance of the Weyl tensor as follows.
According to the equations (\ref{eq:appppB1}) and (\ref{eq:appppB2}), we find that
\begin{align}
	B^{ij}
	=\frac{1}{n-2}(\bar{R}^{ij}-\tilde{\bar{R}}^{ij})
	+\frac{1}{2(n-1)(n-2)}(G^{ij}\bar{R}-\tilde{G}^{ij}\tilde{\bar{R}}).
	\label{eq:appppB3}
\end{align}
By substituting (\ref{eq:appppB3}) into the equation (\ref{eq:appppB4}),
we see that the Weyl tensor $C^{kij}_{l}$ is invariant under the Weyl transformation.

Before closing this section, we remark the local scale transformation of 
the Poisson tensor. 
In the contravariant gravity theory, we consider
 both the Poisson tensor and the metric tensor 
as geometric objects, and we should treat
them equally and independently.
As we have seen, we can consider any local parameter 
$\Omega$ for the Weyl transformation $\tilde{G}^{ij}= e^{2\Omega} G^{ij}$. 
However, we must restrict the local parameter $\omega$ for the transformation 
$\hat{\theta}^{ij}=e^{2\omega}\theta^{ij}$,
because the Poisson condition does not hold unless $d_{\theta}\omega=0$.
Therefore, compared to the Weyl transformation, 
the scale transformation of the Poisson tensor is limited.

%%%%%%%%%%%%%%%%%%%%%%%%%%%%%%%%%%%%
%%%%%%%%%%%%%%%%%%%%%%%%%%%%%%%%%%%%
\section{Gravity on Two-Dimensional Poisson Manifolds}

In the previous section, 
we analyzed general properties of the gravity theory on a Poisson-Riemannian manifold.
For a more concrete discussion, we consider the case of a two-dimensional Poisson-Riemannian manifold 
$(M,g=G^{-1},\theta)$ in detail. 
We reveal the interaction terms between the metric and the Poisson tensor explicitly.

In two dimensions the Poisson tensor has a form
\bea
	\theta^{ij} = \matt[0,\theta,-\theta,0].
\ena
For any function $\theta$ on the manifold, 
the Poisson condition is satisfied
because no three-vector exists on two-dimensional manifolds.
Here, we mention that
the combination
\begin{align}
	&\sqrt{G^{-1}}\theta=\sqrt{g}\theta,	\label{alllph}
\end{align}
is invariant under diffeomorphisms,
since $\theta$ behaves in the same way as $d^{2}x$.
It is convenient to use the combination \eqref{alllph} to discuss interactions 
between the Poisson tensor $\theta^{ij}$ and the metric $g_{ij}=(G^{-1})_{ij}$, 
$g^{ij}=G^{ij}$.

Next, we consider the constraint on the scalar field $\phi$,
which is given by \eqref{Dil}.
On the two-dimensional manifold $M$, the equation reduces to
\begin{align}
	\sqrt{G^{-1}}\theta\partial_{i}\phi=-2\partial_{i}(\sqrt{G^{-1}}\theta).
\end{align}
The general solution for $\phi$ is given by
\begin{align}
	e^{\phi}=C(\sqrt{G^{-1}}\theta)^{-2},\label{eq:kanitama}
\end{align}
where $C$ is an arbitrary constant. 
Therefore the Einstein-Hilbert-type action (\ref{EinHilb}) takes the form
\begin{align}
	S=C\int d^2x\,\frac{1}{\theta^{2}G^{-1}}\sqrt{G^{-1}}\bar{R}.	\label{Mozart}
\end{align}

%%%%%%%%%%%%%%%%%%%%%%%%%%%%%%%%%%%%%%
\paragraph{Field Redefinition}

We start with the action of the form
\begin{align}
	S=\int d^2x\,\frac{1}{\theta^{2}G^{-1}}\sqrt{{G}^{-1}}{\bar{R}},\label{eq:tildestart}
\end{align}
where
the constant $C$ in eq. \eqref{Mozart} is set to unity for simplicity. 
The action can be written as (see appendix \ref{sec:2dd}.1)
\begin{align}
	S
	&=\int d^2x\,\sqrt{g}\left(
	R
	+\frac{2}{\sigma}\nabla^{i}\partial_{i}\sigma
	-\frac{1}{\sigma^2}\partial^{i}\sigma\partial_{i}\sigma
	\right),\label{eq:2daction}
\end{align}
with introducing a scalar degree of freedom defined by
\begin{align}
	\sigma=\sqrt{G^{-1}}\theta ,
\end{align}
where the symbol $\nabla$ represents a covariant derivative 
with respect to the usual Levi-Civita connection.
In this article, we assume that the Poisson manifold does not have a boundary.
Then, we can do an integration by parts freely and obtain the action of the form
\begin{align}
	S&=\int d^{2}x \sqrt{g}\left(R+g^{ij}\partial_{i}\pi\partial_{j}\pi\right) , 
\end{align}
where we have introduced a redefined scalar degree of freedom $\pi$
defined by
\begin{align}
\pi=\log\sigma=\log(\sqrt{g}\theta).
\end{align}
Therefore, we get a free scalar field theory on a two-dimensional Riemannian manifold.

On the other hand,
we could have considered the field redefinition by performing a Weyl transform
applying to \eqref{eq:tildestart}. 
The observation here gives a non-trivial consistency check to the
formula under the Weyl transformation.
The resultant action is
\begin{align}
	S
	=&\int d^2x\,\frac{1}{\theta^{2}G^{-1}}\sqrt{G^{-1}}(\bar{R}+G_{kj}\bar{L}^{kij}_{i}),
		\label{Moz}
\end{align}
with setting the Weyl transformation parameter as
\begin{align}
	\Omega=\log\sigma. \label{eq:Weylpara}
\end{align} 
We must reach the same result as above.
Again, referring to appendix \ref{sec:2dd}.1, the action can be written as
\begin{align}
	S
	&=\int d^2x\,\sqrt{g}\left(
	R
	+\frac{2}{\sigma}\nabla^{i}\partial_{i}\sigma
	-\frac{1}{\sigma^2}\partial^{i}\sigma\partial_{i}\sigma
	+\frac{1}{\sigma^2}G_{kj}L^{kij}_{i}
	\right).	\label{eq:2daction1}
\end{align}
With some manipulations we find (see appendix \ref{sec:2dd}.2)
\begin{align}
	G_{kj}L^{kij}_{i}
	=2\partial^{i}\sigma\partial_{i}\sigma-2\sigma\nabla_{i}\partial^{i}\sigma,\label{eq:goooooooal}
\end{align}
and these terms turn out to give a surface term, together with the invariant measure:
\begin{align}
	\int d^2x\,\frac{\sqrt{g}}{\sigma^2} G_{kj}L^{kij}_{i}
	=-\int d^2x\,\p_i \left(\sqrt{g}\frac{1}{\sigma} G^{ij} \p_j\sigma\right),
\end{align}
which does not change the equation of motion, but modifies the action
equivalently to doing an integration by parts. 
Then, the contravariant Einstein-Hilbert action becomes
\begin{align}
	S
	&=\int d^{2}x \sqrt{g}\left(
	R+\frac{1}{\sigma^2}G^{ij}\p_i \sigma\partial_{j}\sigma
	\right)\\
	&=\int d^{2}x \sqrt{g}\left(R+g^{ij}\partial_{i}\pi\partial_{j}\pi\right) , 
\end{align}
as obtained in the above.

\paragraph{Cosmological Constant}
Under the solution of a scalar field (\ref{eq:kanitama}), the cosmological constant term (\ref{eq:cc}) gives an interaction term of $\sigma$:
\bea
	S_c = 2\Lambda \int d^2 x  e^{\phi}\sqrt{G^{-1}}=2\Lambda \int d^2 x\sqrt{g}\sigma^{-2}.
\ena
In terms of $\pi$, the total action is
\bea
	S+S_{c}=C\int d^{2}x \sqrt{g}\left(R+g^{ij}\partial_{i}\pi\partial_{j}\pi+2\Lambda e^{-2\pi}\right).
\ena

%%%%%%%%%%%%%%%%%%%%%%%%%%%%%%%%%%%%%%%%%
\section{Conclusion and Discussion}%%%%%%%%%%%%%%%%%%%%%%%
%%%%%%%%%%%%%%%%%%%%%%%%%%%%%%%%%%%%%%%%%

In this paper, we studied the gravity theory based on the contravariant Levi-Civita
connection on Poisson manifolds.
In particular, we investigated the Einstein-Hilbert-type action given by 
the scalar curvature constructed from the contravariant Levi-Civita connection.
 We analyzed the property of the invariant measure and 
  found that for the transformation generated by the contravariant Lie derivative $\bar{\mathcal{L}}_\xi$, the invariant measure can be defined 
 only when we impose the condition $\theta(d\xi)=0$ on the parameter one-form $\xi$.

We also discussed about the divergence theorem.
 We found that it is necessary to introduce 
 an additional factor $e^\phi$ 
in the invariant measure 
which satisfies the partial differential equation (\ref{Dil}). 
The reason is that as shown in (\ref{Totd}), the divergence given by the contravariant Levi-Civita connection 
does not become a surface term with the na\"ive invariant measure.
In terms of the $d_{\theta}$-cohomology,
the differential equation (\ref{Dil}) which defines $\phi$ implies that
the one-vector $\bar{\Gamma}^{ij}_{i}\partial_{j}$ is $d_{\theta}$-exact.
Although the vector field $\bar{\Gamma}^{ij}_{i}\partial_{j}$ is
$d_{\theta}$-closed for any Poisson tensor $\theta$, it is not necessarily exact in general.
In this article we assume that $\bar{\Gamma}^{ij}_{i}\partial_{j}$ is exact.
Therefore, the scalar field $\phi$ is always expressed by the metric and the Poisson tensor.
It would be interesting to consider a generalization of the differential equation to 
the gravity theory with nontrivial $d_{\theta}$-cohomology.

We proposed an analog of the Einstein-Hilbert action (\ref{EinHilb}) 
with an invariant measure which is consistent with the divergence theorem. 
Then, we derived the equation of motion for the metric $G^{ij}$. 
This equation is written by the analog of the Einstein tensor
 (\ref{EinT}) in the contravariant gravity theory. Note that we can also add a cosmological term.

The Weyl transformation of 
the Riemann tensor in the contravariant theories is established.
Its behavior under the Weyl transformation is 
very similar to the behavior of the ordinary Riemann tensor 
of the usual Levi-Civita connection.
The scalar field $\phi$ transforms under the Weyl transformation is compatible with the condition (\ref{Dil}).
In addition, we could absorb the dilaton-like coupling
of the scalar field $\phi$ in the contravariant Einstein-Hilbert action, 
i.e. we could move to the Einstein frame in the contravariant gravity theory.

Furthermore, we discussed in detail
the interaction between the Poisson tensor and the metric on the two-dimensional Poisson manifold.
In this case, we solved the differential equation (\ref{Dil}) without
any assumption for the Poisson tensor and the metric,
and we gave an explicit form of the contravariant Einstein-Hilbert action.
In addition, we showed that the action can be expressed by the ordinary Einstein-Hilbert action 
coupled to a free scalar field $\pi$.
We also discussed the interaction term induced
by the cosmological constant term.
The interaction turned out to be described by
an exponential potential of the scalar field $\pi$ in the usual Einstein theory.

There are some interesting future directions to study the contravariant gravity theory.
In general, we can add all possible terms which are compatible with contravariant gravity,
for instance a kinetic term of the Poisson tensor.
It would also be natural to ask whether one can move on to noncommutative spaces, 
applying the quantization of the Poisson structure in \cite{KM}.
In this sense, as we mentioned in the introduction, our theory may 
be related to gravity theories on noncommutative spaces 
discussed in the purely mathematical as well as  
in the string/M-theoretical literature, such as \cite{BGGK,HR,SH,Hitchin,Connes,JW,J,KS,P,A,KA}.

Recently, a notion of  emergent geometry has been discussed
within matrix theoretic and noncommutative geometric frameworks \cite{Y,Stein1,Stein2,Stein,I}.
In those considerations, the Poisson tensor plays a role as significant as the metric.
Recalling that our theory handles the Poisson tensor on an equal footing with the metric, it would be interesting to apply our framework to geometries 
arising from matrix models of superstring theory and M-theory,
especially those equipped with the K\"ahler structure \cite{Iq,Y1,Y2,IMM}.

It would be also interesting to use our framework for the description of effective theories of  superstring theory.
Actually, our original motivation to introduce the gravity with the (quasi-) Poisson structure
and, as it turned out, with the contravariant Levi-Civita connection
was to describe aspects of T-duality such as non-geometric background fields,
e.g. R-flux \cite{AMW,AMSW,Andriot:2011uh,Blumenhagen:2012nt,Andriot:2012wx,Andriot:2012an,Andriot:2013xca,Andriot:20141,Andriot:20142}.
In these references, the same connection is utilized to formulate the non-geometric fluxes.
On the other hand, the non-geometric background with Poisson structure is also analyzed
by using the supergeometric method in a recent study \cite{BHIW}.
It is also interesting to discuss the gravity theory with Poisson structure from the supergeometric point of view \cite{Schwarz1,Schwarz2,AKSZ,Ikeda,CHIKW}.

Finally, one can consider the contravariant gravity theory in odd-dimensional spacetime
where the Poisson tensor must be degenerate. Note that we did not need  the 
non-degeneracy of the Poisson tensor in our analysis.
The degeneracy would clarify
the difference between the gravity theory based on the contravariant Levi-Civita connection
and that of the usual Levi-Civita connection,
since
the degeneracy of the Poisson tensor gives rise to 
a degeneracy of the contravariant Lie derivatives,
which then might fail to generate full diffeomorphisms.
This defect of diffeomorphisms would give new aspects in the gravity theory. The dimensional 
reduction has a similar effect.
Along this line,  a dimensional reduction
has been already discussed 
in part on the Courant algebroid level \cite{Q}.
As a related study, the (generalized) Einstein-Hilbert
actions and their dimensional reductions have been also discussed
from a viewpoint based on the Courant/Leibniz algebroid by \cite{JV1,JV2}.
Besides, in odd-dimensional spaces there are the contact structure, 
the Jacobi structure and the Nambu-Poisson structure,
as specific structures related to the Poisson structure.
It is an open question how the gravity theory coupled to such structures could be realized.
The investigation of these structures is a near future project and we hope that 
this will shed some light on the properties of odd-dimensional noncommutative and nonassociative spaces.

%%%%%%%%%%%%%%%%
%%%%%%%%%%%%%%%%

\section*{Acknowledgments}

The authors would like to give thanks to Tsuguhiko Asakawa for helpful comments and
important contributions at the earlier stage of this work, 
Ursula Carow-Watamura for careful reading and improving our manuscript,
Taiki Bessho, Marc Andre Heller,
Noriaki Ikeda, Goro Ishiki, Branislav Jur\v{c}o, Shinpei Kobayashi, Takaki Matsumoto,
Yuta Sekiguchi and Thomas Strobl
for fruitful discussions.
YK is supported by Tohoku University Division 
for Interdisciplinary Advanced Research and Education (DIARE).
HM is supported in part by the Iwanami Fujukai Foundation.

%%%%%%%%%%%%%%%%%%%%%%%%%%%%%%%
%%%%%%%%%%%%%%%%%%%%%%%%%%%%%%%

\appendix

%%%%%%%%%%%%%%%%%%%%%%%%%%%%%%%%%%%%
\section{Note on Riemannian Geometry on Poisson Manifolds \label{ApenDA}}%%%%%%%

In this section we give a brief note on how one can build the notion of Riemannian geometry 
on Poisson manifolds.
First, in order to fix our conventions, we give a quick review on
a Lie algebroid of one-forms induced by a Poisson structure on Poisson manifolds.
Precise statements and detailed arguments can be found in \cite{AMW,AMSW}
and also in \cite{fernandes2000,Boucetta:2011ofa2,Boucetta:2011ofa3,KSmodular,Hawkins:2002rf}.
Based on the Lie algebroid,
we define a contravariant affine connection, its torsion and curvature tensors.

%%%%%%%%%%%%%%%%%%%%%%%%%%
\paragraph{Lie Algebroid on Poisson Manifolds}

Let $M$ be an $n$-dimensional Poisson manifold equipped with a Poisson tensor
$\theta \in \Gamma(\wedge^2 T M)$,
which satisfies the condition $[\theta,\theta]_S=0$.
Here $[\cdot,\cdot]_S$ stands for the Schouten-Nijenhuis bracket.
A Lie algebroid on the manifold $M$ is defined 
by a triple $(T^*M, \, \theta, \, [\cdot,\cdot]_\theta)$:
$T^*M$ is the cotangent bundle over $M$;
the anchor map $\theta :T^*M \to T M$ is given by the Poisson tensor through
$\theta(\xi)=\bar{\iota}_\xi \theta$ for $\xi \in \Gamma(T^*M)$;
and the Lie bracket is given by 
\bea
	[\xi,\eta]_\theta ={\cal L}_{\theta (\xi)}\eta-i_{\theta (\eta)}d\xi ,
\ena
called the Koszul bracket.

An exterior derivative $d_\theta =[\theta, \cdot]_S $,
an interior product $\bar{\iota}_\zeta$
and a Lie derivative $\bar{\cal L}_{\zeta}=\{ d_\theta ,\bar{\iota}_\zeta\}$ 
with respect to a one-form $\zeta \in \Gamma(T^*M)$
acting on polyvectors $\Gamma(\wedge^\bullet T M)$ are defined. 
Due to the Poisson condition,
the nilpotency of the exterior derivative, $d^2_\theta=0$, 
is guaranteed,
so that the exterior derivative $d_\theta$ defines the $d_\theta$-cohomology.
These operations satisfy the Cartan relations 
\bea
	\{ \bar{\iota}_\xi, \bar{\iota}_\eta\}=0, \quad
	\{ d_\theta ,\bar{\iota}_\xi\}=\bar{\cal L}_\xi, \quad
	[\bar{\cal L}_\xi, \bar{\iota}_\eta ]=\bar{\iota}_{[\xi,\eta]_\theta},\quad
	[\bar{\cal L}_\xi, \bar{\cal L}_\eta ]=\bar{\cal L}_{[\xi,\eta]_\theta},\quad
	[d_\theta,\bar{\cal L}_\xi ]=0.
\label{A Cartan}
\ena

%%%%%%%%%%%%%%%%%%%%%%%%%%%%%%%%%%%%
\paragraph{Contravariant Affine Connection\label{CaffineC}}%%%%%%%%%%

Let us introduce a notion of  contravariant affine connection $\bar\nabla$ on 
the Lie algebroid $\Gamma(T^*M)$. It is defined as a map
satisfying
\bea
	\bar\nabla_{f\xi} \eta =f \bar\nabla_\xi \eta, 
	~~\bar\nabla_\xi (f\eta)=(\bar\cL_{\xi} f)\eta+f \bar\nabla_{\xi} \eta , \label{affine}
\ena
for any $1$-forms $\xi$, $\eta$ and function $f$.
Thus, the covariant derivatives are global objects and
independent of our choice of local coordinates.

On some local patch $\{x^i\}$, the connection above is specified by the coefficients
\bea
	\bar\nabla_{dx^i}dx^j = \bar{\Gamma}^{ij}_k dx^k, \label{25}
\ena
for one-form basis $\{dx^i\}$.
On another local patch, say $\{{x'}^i\}$, we have another set of basis 
of one-forms $\{d{x'}^i\}$. For this basis, we introduce
the coefficients $\bar{\Gamma}'$ as
\bea
	\bar{\nabla}_{d{x'}^a} d{x'}^b
			&\equiv {(\bar{\Gamma}')}^{ab}_c d{x'}^c.
\ena
On the intersection of the two local patches $\{{x}^i\}$ and $\{{x'}^i\}$,
on the other, we also have
\bea
	\bar{\nabla}_{d{x'}^a} d{x'}^b
			&= \bar{\nabla}_{\frac{\p {x'}^a}{\p x^i}dx^i} \bigg(\frac{\p {x'}^b}{\p x^j} dx^j\bigg)
			=\frac{\p{x'}^a}{\p x^i}\bigg(\ \theta^{ij}\frac{\p^2 {x'}^b}{\p x^j \p x^k}
				+\frac{\p {x'}^b}{\p x^j}\bar{\Gamma}^{ij}_k\bigg)dx^k,
\ena
with a use of \eqref{affine}  and \eqref{25}.
Then we can extract the behavior of the coefficients under coordinate transformations.
As usual, the coefficients do not behave as a tensor:
\bea
	{(\bar{\Gamma}')}^{ab}_c 
	= \frac{\p x^k}{\p {x'}^c}
		\frac{\p {x'}^a}{\p x^i} \theta^{ij}\frac{\p^2 {x'}^b}{\p x^j \p x^k}
		+\frac{\p {x'}^a}{\p x^i}\frac{\p {x'}^b}{\p x^j}\frac{\p x^k}{\p {x'}^c}\bar{\Gamma}^{ij}_k.
		\label{trpG}
\ena
This property \eqref{trpG} implies that
the subtraction $\bar{\Gamma}^{(1)}-\bar{\Gamma}^{(2)}$ of
any two connections $\bar{\Gamma}^{(1)}$ and $\bar{\Gamma}^{(2)}$
behaves as a tensor of type $(2,1)$,
as long as the same Poisson tensor is utilized.
And also the trace of the connection coefficients
$\{\bar{\Gamma}^{ab}_a\}$ defines a tensor of type $(1,0)$, i.e. a vector field.

%%%%%%%%%%%%%%%%%%%%%%%%%%%%%%%%%%%%
\paragraph{Torsion and Curvature}%%%%%%%%%%%%%%%%%%%

The torsion of a contravariant affine connection $\bar \nabla$ is defined 
by
\bea
	\bar{T}(\xi,\eta)=\bar\nabla_{\xi} \eta 
				-\bar\nabla_{\eta} \xi - [\xi,\eta]_\theta.  \label{2.15}
\ena
$\bar{T}$ turns out to be a tensor of type $(2,1)$, since it satisfies
$\bar{T}(f\xi,\eta)=f\bar{T}(\xi,\eta)=\bar{T}(\xi,f\eta)$.

The curvature of a contravariant affine connection $\bar \nabla$ is defined by
\bea
	\bar{R}(\xi,\eta)\zeta=
	(\bar{\nabla}_\xi \bar{\nabla}_\eta -\bar{\nabla}_\eta \bar{\nabla}_\xi 
		- \bar \nabla_{[\xi,\eta]_\theta} )\zeta.
\ena
It is easily shown that, for any functions $f,g$ and $h$, it satisfies
$\bar{R}(f\xi,g\eta)(h\zeta)=fgh\bar{R}(\xi,\eta)\zeta$.

The curvature, together with the torsion tensor $\bar{T}$, satisfies
the following Bianchi identities
\bea
	&\mathfrak{S}\{\bar{R}(\xi,\eta) \zeta\} 
	=\mathfrak{S}\{(\bar{\nabla}_{\zeta}\bar{T})(\xi,\eta)
	+\bar{T}(\bar{T}(\xi,\eta),\zeta)\},\label{bi1}\\
	&\mathfrak{S}\{(\bar{\nabla}_\zeta \bar{R})(\xi,\eta)
	+\bar{R}(\bar{T}(\xi,\eta),\zeta)\}=0,\label{bi2}
\ena
where $\mathfrak{S}$ denotes the cyclic sum over $\xi$, $\eta$ and $\zeta$.

%%%%%%%%%%%%%%%%%%%%%%%%%%%%%%%
\section{Relation between (Contravariant) Levi-Civita Connections}\label{cova-contra}%%%%%

The contravariant Levi-Civita connection is related to the usual Levi-Civita connection by
\bea 
	\bar{\Gamma}^{ij}_k
	=\Gamma^j_{mk} \theta^{mi}    +K^{ij}_k,	\label{covariant-contravariant}
\ena
where $K^{ij}_k$ is understood as a contravariant version of contorsion tensor
\bea
	& {K_k}^{ij}=G_{kl}K^{lij},
		~~K^{kij} =\frac{1}{2}\left(\nabla^k \theta^{ij} -\nabla^i \theta^{jk}
		+\nabla^j \theta^{ki}\right) .
\ena
Here $\nabla$ denotes the ordinary usual Levi-Civita connection.
The raising and lowering of the indices are done by the metric $G$ and $G^{-1}$ as usual, 
e.g. $\nabla^i=G^{ij}\nabla_j$.
The contravariant Riemann tensor in terms of the usual Levi-Civita connection and the contorsion tensor reads
\bea
	&\bar{R}^{kij}_{l}
	= \theta^{im} \theta^{nj} {\sf R}^k_{lmn}
		-  (\nabla_n \theta^{ij}) K^{nk}_l
 		+ \theta^{nj} \nabla_n K^{ik}_l 
		 -\theta^{ni} \nabla_n K^{jk}_l
		+K^{jk}_m  K^{im}_l 
		-  K^{ik}_mK^{jm}_l  , \label{riempois}
\ena
where ${\sf R}^k_{lmn}$ is the ordinary Riemann tensor made out of the metric $G_{ij}$.
By using the expression \eqref{riempois}, the contravariant Ricci tensor and the scalar curvature are also written as 
\bea
	&\bar{R}^{kj}
  	=\theta^{lm} \theta^{nj} {\sf R}^k_{lmn}
		-  (\nabla_n \theta^{lj}) K^{nk}_l
 		+ \theta^{nj} \nabla_n \nabla_l \theta^{lk} 
		 -\theta^{nl} \nabla_n K^{jk}_l
		+K^{jk}_m  \nabla_l \theta^{lm} 
		-   K^{lk}_m K^{jm}_l   ,\nn
	&\bar{R}
	= \theta^{lm} \theta^{nj} {\sf R}_{jlmn}
 		+ 2\theta_{nm} \nabla^n \nabla_l \theta^{lm} 
				- \nabla^n \theta_{nm}  \nabla_l \theta^{lm} ,	\label{ravel}
\ena
where $\theta_{ij}=G_{ik}G_{jl}\theta^{kl}$.

%%%%%%%%%%%%%%%%%%%%%%%%%%%%%%%%%%%%%%
\section{Computations on Two-dimensional Manifolds}\label{sec:2dd}

Some straightforward but lengthy computations are presented.

%%%%%%%%%%%%%%%%%%%%%%%%%%%%%%%%%%%%%%
\subsection{Calculation of Ricci Tensor}

Since the Riemann tensor $\bar{R}^{ijk}_{l}$ have the same structure as $R^{i}_{jkl}$, 
the only independent component of the Riemann tensor is
\begin{align}
	\bar{R}^{1212}=\frac{1}{2}(\text{det}G) \bar{R},
\end{align}
i.e.
\begin{align}
	\bar{R}^{ijkl}=\frac{1}{2}(G^{ik}G^{jl}-G^{il}G^{jk}) \bar{R}.
\end{align}
This equation implies that the vacuum Einstein equation \eqref{eom1} is automatically satisfied.

The Ricci scalar in the usual covariant language is given by \eqref{ravel}, i.e.
\begin{align}
	\bar{R}=\theta^{jk}\theta^{li}R_{ijkl}+2\theta_{kj}\nabla^{k}\nabla_{i}\theta^{ij}
	-\nabla^{k}	\theta_{kj}\nabla_{i}\theta^{ij},\label{eq:panda}
\end{align}
where $R_{jlmn}$ is the Riemann tensor made out of $g_{ij}$, $\nabla$ denotes
the usual Levi-Civita connection and the indices are lowered by the Riemannian metric $g_{ij}$:
\begin{align}
	\theta_{12}&=G_{1i} G_{2j}\theta^{ij}
	=(G_{11} G_{22}-G_{12} G_{21})\theta^{12}
	= (\det G^{-1})\theta^{12},
\end{align}
which implies
\bea
	\theta_{ij}&=\epsilon_{ij}(\det g){\theta}
			=\epsilon_{ij} \sqrt{g} \sigma. 
\ena
Here we have introduced $\sigma=\sqrt{g}\theta.$

In the following we demonstrate explicit computations of
each terms in (\ref{eq:panda}). We can see immediately that the first term is
\begin{align}
	\theta^{jk}\theta^{li}R_{ijkl}
	=&\theta^2(\text{det}g) R.
\end{align}
To calculate the second and third terms, we consider a covariant derivative 
acting on the Poisson tensor:
\begin{align}
	\nabla_{k}\theta^{ij}
	=&\partial_{k}\theta^{ij}+\Gamma^{i}_{kl}\theta^{lj}+\Gamma^{j}_{kl}\theta^{il}\nn
	=&(\partial_{k}\sigma)\frac{1}{\sqrt{g}}\epsilon^{ij}+\frac{1}{\sqrt{g}}\left(
	-\frac{1}{\sqrt{g}}\partial_{k}\sqrt{g}\epsilon^{ij}+\Gamma^{i}_{km}\epsilon^{mj}+
	\Gamma^{j}_{km}\epsilon^{im}
	\right)	.
\end{align}
Since $\Gamma^{i}_{kl}\epsilon^{lj}+\Gamma^{j}_{kl}\epsilon^{il}$ 
is antisymmetric under an exchange between $i$ and $j$,
we have the relation
\begin{align}
	\Gamma^{i}_{kl}\epsilon^{lj}+\Gamma^{j}_{kl}\epsilon^{il}
	=\epsilon^{ij}\Gamma^{l}_{lk}=\epsilon^{ij}\frac{1}{\sqrt{g}}\partial_{k}\sqrt{g}.
\end{align}
Therefore we find the covariant derivative on the Poisson tensor becomes
\begin{align}
	\nabla_{k}\theta^{ij} =(\partial_{k}\sigma)\frac{1}{\sqrt{g}}\epsilon^{ij}.
\end{align}
Utilizing this, we have
\begin{align}
	\nabla_{k}\theta_{ij}
	=&G_{im}G_{jn}\nabla_{k}\theta^{mn}
	=\epsilon_{ij} {\sqrt{g}} \partial_{k}\sigma .
\end{align}
Using above relations, we find the second and third terms in (\ref{eq:panda}) become
\begin{align}
	\theta_{ij}\nabla^{i}\nabla_{k}\theta^{kj}
	=&{\sigma}\nabla^{i}\partial_{i}\sigma,\nonumber\\
\nabla^{i}\theta_{ij}\nabla_{k}\theta^{kj}
	=& \partial^{i}\sigma\partial_{i}\sigma .
\end{align}
Therefore, we get an explicit form of the Ricci scalar in terms of $\sigma$:
\begin{align}
	\bar{R}=\sigma^{2}R
	+2{\sigma}\nabla^{i}\partial_{i}\sigma
	-\partial^{i}\sigma\partial_{i}\sigma.
\end{align}

%%%%%%%%%%%%%%%%%%%%%%%%%%%%%%%%%%%%%%
\subsection{Calculation of Weyl Transformation} 

In the following, we calculate $G_{kj}\bar{L}^{kij}_{i}$ explicitly. 
On a two-dimensional manifold we find
\bea
	G_{kj}\bar{L}^{kij}_{i}
	&=G_{kj}(\delta^{j}_{i}B^{ik}-2B^{jk}-G^{jk}G_{im}B^{im}+\delta^{j}_{m} B^{mk})=-2G_{ij}B^{ij},
\ena
where
\bea
	G_{ij}B^{ij}&=
	G_{ik}\left(
	\bar{\nabla}_{dx^{i}}(\theta^{kj}\partial_{j}\Omega)
	-(\theta^{ij}\partial_{j}\Omega)(\theta^{kl}\partial_{l}\Omega)
	+\frac{1}{2}G^{ik}G_{jl}(\theta^{jm}\partial_{m}\Omega)(\theta^{ln}\partial_{n}\Omega)\right)\nn
	&=\bar{\nabla}_{dx^{i}}(G_{ik}\theta^{kj}\partial_{j}\Omega).
\ena
Thus we see, with a use of \eqref{Beeth},
\bea
	\frac{1}{2}e^\phi\sqrt{g}G_{kj}\bar{L}^{kij}_{i}
	&=-e^\phi\sqrt{g}\bar{\nabla}_{dx^{i}}(G_{ik}\theta^{kj}\partial_{j}\Omega)\\
	&=- \p_i (e^\phi\sqrt{g}G_{ik}\theta^{kj}\partial_{j}\Omega).
\ena 
Hence an additional term in \eqref{Moz} induced by a Weyl transform
is just a surface term for any choice of $\Omega$.
Furthermore, in the case (\ref{eq:Weylpara}) we have
\bea
	G_{ij}B^{ij}&=\theta^{il}\p_l \left(\frac{G_{ik}}{\sigma}\theta^{kj}\p_j\sigma\right)
	+ \frac{1}{\sqrt{G^{-1}}}\p_i(\sqrt{G^{-1}}\theta^{il}) G_{lk} \theta^{kj} \frac{\p_j\sigma}{\sigma}.
\ena
Here we utilized the fact $\bar{\nabla}_{dx^k}G_{ij}=0$ 
and the formula \eqref{debussy}.
Substituting $\theta^{ij}=\frac{1}{\sqrt{G^{-1}}}\sigma\epsilon^{ij}$ 
and using $\epsilon^{ik}\epsilon^{jl}G_{kl}=G^{-1}G^{ij}$,
we find that
\bea
	G_{kj}\bar{L}^{kij}_{i}
	&=   2G^{ij} \p_i \sigma \p_j\sigma -2\sigma G^{ij}   \nabla_i \p_j\sigma ,
\ena
and then
\bea
	\frac{1}{2}e^\phi\sqrt{g}G_{kj}\bar{L}^{kij}_{i}
	&=\sqrt{g}\frac{1}{\sigma^2}( G^{ij} \p_i \sigma \p_j\sigma -\sigma G^{ij}   \nabla_i \p_j\sigma).
\ena

%%%%%%%%%%%%%%%%%%%%
%%%%%%%%%%%%%%%%%%%%

\providecommand{\href}[2]{#2}\begingroup\raggedright
\endgroup

\end{document}